\documentclass{aa}
\usepackage{graphics}
\begin{document}

\title{Magnetic fields and ionized gas in nearby late type galaxies$^*$
}

\author {
K.T. Chy\.zy\inst{1}
\and D.J. Bomans\inst{2}
\and M. Krause\inst{3}
\and R. Beck\inst{3}
\and M. Soida\inst{1}
\and M. Urbanik\inst{1}}
\institute{Astronomical Observatory, Jagiellonian
University, ul. Orla 171, 30-244 Krak\'ow, Poland
\and Astronomisches Institut, Ruhr-Universit\"at-Bochum,
44780 Bochum, Germany
\and Max-Planck-Institut f\"ur Radioastronomie, Auf dem
H\"ugel 69, 53121 Bonn, Germany}

\offprints{K.T. Chy\.zy}
\mail{chris@oa.uj.edu.pl\\
$^*$Based on the observations with the 100-m telescope at Effelsberg 
operated by the Max-Planck-Institut f\"ur Radioastronomie (MPIfR) on behalf 
of the  Max-Planck-Gesellschaft.}
\date{Received date/ Accepted date}

\titlerunning{Magnetic fields and ionized gas in late type galaxies}
\authorrunning{K.T. Chy\.zy et al.}

\abstract
{}
{In order to analyze the importance of the star formation rate in generating
and amplifying magnetic fields in the interstellar medium we perform a deep
continuum polarization study of three angularly large, late type spiral
galaxies.}
{ We obtained deep total power and polarization maps at
4.85\,GHz of NGC 4236, NGC 4656 and IC 2574
using the 100-m  Effelsberg radio telescope. This
was accompanied by imaging in the H$\alpha$ line. We also observed
these objects at 1.4\,GHz to  obtain their integrated fluxes
at this frequency and to determine their radio spectra.}
{ All galaxies were found to  possess weak but detectable total power 
emission at 4.85\,GHz, coincident with regions of recent star formation as
traced by bright H$\alpha$ regions. 
The surface brightness of the radio-strongest object of our sample 
(NGC 4656) is comparable to the radio-weakest objects
in a sample of more than 50 normally star-forming spiral galaxies for
which measurements at 4.8\,GHz with the Effelsberg radio
telescope are available. The surface brightness of the two other objects 
is even three times smaller. The fractional polarization of the 2 galaxies 
of our sample is less than 2\%, significantly lower than for spiral galaxies of 
intermediate types, suggesting that the magnetic fields are not only weaker,
but also less ordered than in spiral galaxies.
The radio spectra of galaxies in our small sample are indicative of a 
substantial fraction of thermal emission, with a higher thermal fraction
than in spirals with high star formation rates (SFR), while the nonthermal 
emission in our sample is relatively weak compared to spiral galaxies.
We propose an equipartition model where the nonthermal emission increases 
$\propto SFR^{~\approx 1.4}$ and the ratio of nonthermal to thermal 
emission increases $\propto SFR^{~\approx 0.4}$.
The objects of our sample still follow the radio-FIR correlation of
surface brightness of the total emission, extending it towards the 
lowest values measured so far.
}
{}


\maketitle

\section{Introduction}

The generation of galactic magnetic fields requires strong dynamo 
action  (Beck et al. \cite{beck96}) driven by Coriolis forces
caused by rapid  disk  rotation. As suggested by Chy\.zy et
al. (\cite{4449}) rotation alone is not the only agent in
generating the  large-scale regular and random magnetic fields.
In the classical dynamo theory as well as in more recent
dynamo  concepts (e.g. Blackman \cite{black98}, Schekochihin et
al. \cite{sch04}) the energy input from interstellar turbulence
is  required. Turbulence itself is believed to be powered by
star-forming  processes. A promising mechanism to generate
quickly strong magnetic fields, even  with weak Coriolis forces,
is the dynamo mechanism based on the Parker instability (Kowal et al.
\cite{kowal}, Hanasz et al. \cite{hanasz04}, Moss et al.
\cite{moss}). This process depends directly on
the formation of massive stars, because the instabilities grow due to
the pressure of cosmic rays (CR) produced by supernovae.
This model explains very well the stronger magnetic fields in
rapidly star-forming  spirals. On the other hand, this may mean
that in galaxies forming stars very slowly the instabilities cannot
produce widespread magnetic fields.

The above relation seems to be supported by observations of the low
surface brightness irregular galaxy NGC\,6822 by Chy\.zy et al.
(\cite{chris_ic}). This very slowly star-forming galaxy shows
only weak signs of global magnetic fields. Until now no
detailed radio continuum study of such objects was made.  In
particular it is not  known whether at the low end  of the
surface brightness distribution the magnetic field strength
decreases continuously in some proportion to the star formation
rate or drops  suddenly to zero below a certain threshold of
star-forming activity.

In a search for such a threshold we undertook a study of three
late type galaxies -- NGC\,4236, NGC\,4656, IC\,2574 -- showing
rotational and other properties intermediate  between spirals and
irregulars (Tab.~\ref{prop}). We have chosen galaxies of
substantial  inclination to ensure a long pathlength through the
disk and hence a higher observed surface brightness.  They
all have rather low surface brightnesses with different
star-forming properties. While NGC\,4236 has a star-forming
activity symmetrically distributed over the disk,  NGC\,4656 shows
almost all star formation concentrated in one of the disk halves. 
The third galaxy, IC\,2574, has  a very low star formation level,
restricted to one  region in the disk outskirts. NGC\,4236 and
IC\,2574 belong to the M81 group and  both show similar
color-magnitude diagrams (Karachentsev et al. \cite{kara}). 
IC\,2574 seems to be optically very diffuse and does not reveal
any obvious nucleus  (Ho et al. \cite{ho}). NGC\,4656 belongs
to the group of NGC\,4631, to which the distance is only 47\,kpc and in
which direction some \ion{H}{i} filaments are observed  (Rand 
\cite{rand}).

The galaxies are angularly large, with D$_{25}$ ranging
from 13\arcmin \,  to  23\arcmin \, (see Tab. 1). Their
expected radio surface brightness is too low to be detectable at
centimeter wavelengths with the VLA even with the most compact configuration.
On the other  hand, they are large enough to be resolved at
4.85\,GHz with $2\farcm5$ HPBW resolution of the Effelsberg telescope 
which enables to identify large-scale star formation
distributions. We decided to use this instrument and
frequency for our study. To obtain detailed information
on their star-forming   properties, the galaxies were mapped in
the H$\alpha$ line with the 1-m telescope  at Mt. Laguna
Observatory. This kind  of study also brings information about
the ionized gas properties  in slowly  star-forming, late-type
spirals and especially about the thermal free-free emission. To 
make another  estimate of their thermal emission from the radio
spectrum we  measured also their integrated flux densities at
1.4\,GHz with the Effelsberg  telescope. Confusion by unresolved
background sources is severe at this frequency, but high
resolution data are available at 1.49\,GHz (NVSS, FIRST, Condon
\cite{condon}),  enabling the subtraction of point sources 
unrelated to galaxies from our measurements.

\begin{table*}
 \begin{center}
 \caption[]{Basic properties of NGC\,4236 NGC\,4656 and IC\,2574 (mainly from 
the LEDA database)}
 \begin{tabular}{lllll}
 \hline
 & \object{NGC 4236} & \object{NGC 4656} & \object{IC 2574}&\\
 \hline
 R.A.$_{2000}$ & $\rm 12^h16^m43\fs1$ & $\rm 12^h43^m58\fs2$ &
 $10^h28^m21\fs5$&\\
 Dec.$_{2000}$ & $+69\degr 27\arcmin51\arcsec$ &
 $+32\degr 10\arcmin 14\arcsec$ & $+68\degr24\arcmin 41\arcsec$ &\\
 Inclination$^{a}$ & 75$\degr$$^{b}$ & 83$\degr$$^{c}$ & 77$\degr$$^{d}$&\\
 Position Angle & 162$\degr$ & $35\degr$ &$50\degr$&\\
 Morphol. Type & SBdm & SBm & SABm&\\
 Optical diameter D$_{25}$  & 23\farcm 4 & 13\farcm 5 & 13\farcm 5&\\
 Distance [Mpc]         & 4.45$^{e}$ &  7.5$^{c}$    & 4.02$^{e}$\\
 B-magnitude          & $10\fm 06$ & $11\fm 35$ & $10\fm 80$\\
 Absolute B-magnitude & $-18\fm 01$ & $-20\fm 57$ & $-16\fm 95$\\
 Mean surface brightness [mag/arcsec$^2$] & $24\fm 61$ & $24\fm 27$ & $24\fm 52$ \\
 \ion{H}{i} mass [M$_\odot$]$^{c}$          & $1.5\times 10^9$ & $5\times 10^9$  & $6.7\times 10^8$  \\
 Rotation velocity (from \ion{H}{i} [km/s]) & $\approx 85$$^{b}$ & $\approx 75$$^{c}$  & 67$^{d}$ \\
 \hline \\
$^{a}$ $0\degr$ = face-on\\
$^{b}$ Honma \& Sofue (\cite{honma})\\
$^{c}$ Rand (\cite{rand})\\
$^{d}$ Martimbeau et al. (\cite{martimbeau})\\
$^{e}$ Karachentsev et al. (\cite{kara})\\

\label{prop}
\end{tabular} 
\end{center} 
\end{table*}

\section{Observations and data reduction}

\subsection{H$\alpha$ images}

\begin{figure}
\resizebox{\hsize}{!}{\includegraphics{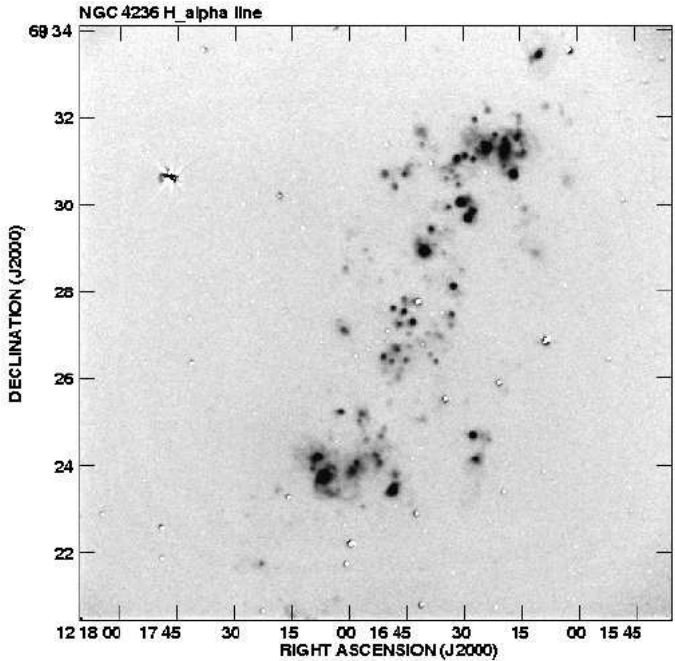}}
 \caption{
The H$\alpha$ image of NGC\,4236 with the optical continuum subtracted
}
\label{4236hal}
\end{figure}

The optical data for our galaxies were taken with the 1-m
telescope at Mt. Laguna  Observatory (operated by the San 
Diego State University, California) using a Loral/Lesser
$2048^2$ pixel CCD. The data reduction techniques were the same
as described by Chy\.zy et al. (\cite{chris_ic}) and performed 
in the standard way using the IRAF package. We applied 
small kernel Gaussian filters to both the H$\alpha$ 
and the R image to match the slightly different point
spread functions. After this step the continuum  subtraction was
performed, as described in e.g. Bomans et al. (\cite{boman97}). 
The maps were calibrated to the absolute scale and corrected for
the \ion{N}{ii}  contribution.  

\subsection{Radio continuum}

The total power and polarization observations were performed at
4.85\,GHz with 2\farcm 5~HPBW resolution using the two-horn system
in the secondary focus of the Effelsberg 100-m MPIfR telescope
(Schmidt et al. \cite{schmidt}). The telescope pointing was 
corrected by making cross-scans of bright point sources at
time intervals of about 1.5 hour. The flux density was
calibrated on the highly polarized  source 3C286. A total power
flux density at 4.85\,GHz of 7.47\,Jy was adopted using the
formulae by Baars at al. (\cite{baars}). The polarized flux
density was calibrated using the same factors as for total power,
yielding a degree of  polarization of 10.5\% for 3C286, which is
in good agreement with other published values (Tabara \&~Inoue
\cite{tabara}). At 4.85\,GHz we observed 
NGC\,4236, NGC\,4656 and IC\,2574  in the
azimuth-elevation frame with field sizes of $40\arcmin\times30\arcmin$,
$46\arcmin\times30\arcmin$ and $46\arcmin\times36\arcmin$, respectively.
The pixel size and separation between scans of 
1\arcmin\ fulfill the necessary (Nyquist) sampling requirement. 
The scaning velocities were 30\arcmin/min for NGC\,4236 and NGC\,4656, 
and 50\arcmin/min for IC\,2574.
We used the NOD2 data reduction package (Haslam
\cite{haslam}). By combining the information from the two
horns, using the ``software beam switching'' technique (Morsi
\&~Reich \cite{morsi}) followed by restoration of  total
intensities (Emerson et al. \cite{emerson}), we obtained I, Q and
U maps  for each coverage of a given galaxy. 
Those channel maps with an excess signal along the scanning 
direction were removed from further analysis, 
resulting in 47 total power and 36 polarized channel maps for NGC\,4236, 
19 and 19 maps for NGC\,4656, and 20 and 19 maps for IC\,2574,
respectively.
All good maps were then combined  using the spatial-frequency weighting method
(Emerson \&~Gr\"ave \cite{emgra}), followed by a digital
filtering process, that removed the spatial  frequencies
corresponding to noisy structures smaller than the telescope
beam.  Finally the I, Q and U images were combined into the maps
of total power,  polarized intensity, polarization degree and
polarization position angles. The r.m.s. noise levels in the
final maps of total intensity are 0.4\,mJy/b.a., 0.7\,mJy/b.a.
and 0.4\,mJy/b.a. for NGC\,4236, NGC\,4656 and IC\,2574,
respectively. The corresponding noise levels in polarized
intensity are 0.04\,mJy/b.a., 0.07\,mJy/b.a. and 0.08\,mJy/b.a.

\begin{figure}
\resizebox{\hsize}{!}{\includegraphics{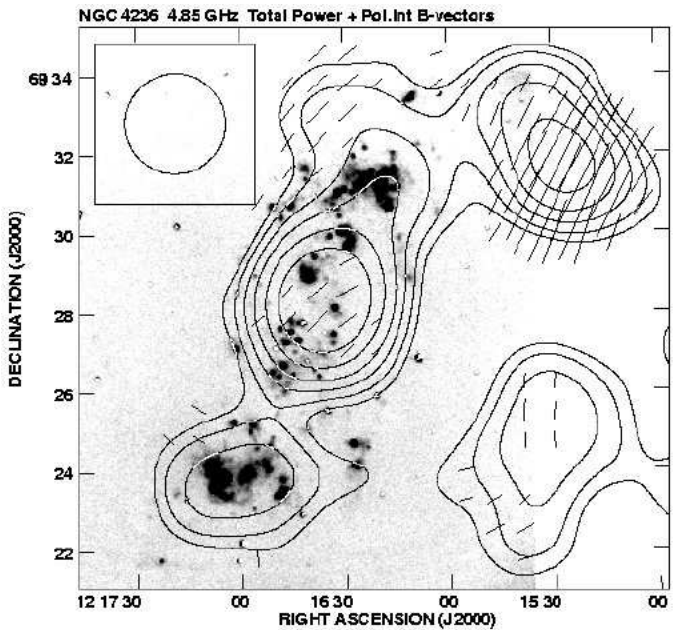}}
\caption{
Total power contours and B-vectors of polarized intensity of NGC\,4236 at 
4.85\,GHz superimposed onto the H$\alpha$ image. The contour levels are (3, 5, 
8, 12, 18, 26) $\times$ 0.3\,mJy/b.a. A vector of length of 1$'$ corresponds to 
a polarized intensity of 0.25\,mJy/b.a.  The map resolution is 2\farcm 5~HPBW.
}
\label{4236tp}
\end{figure}

\begin{figure}
\resizebox{\hsize}{!}{\includegraphics{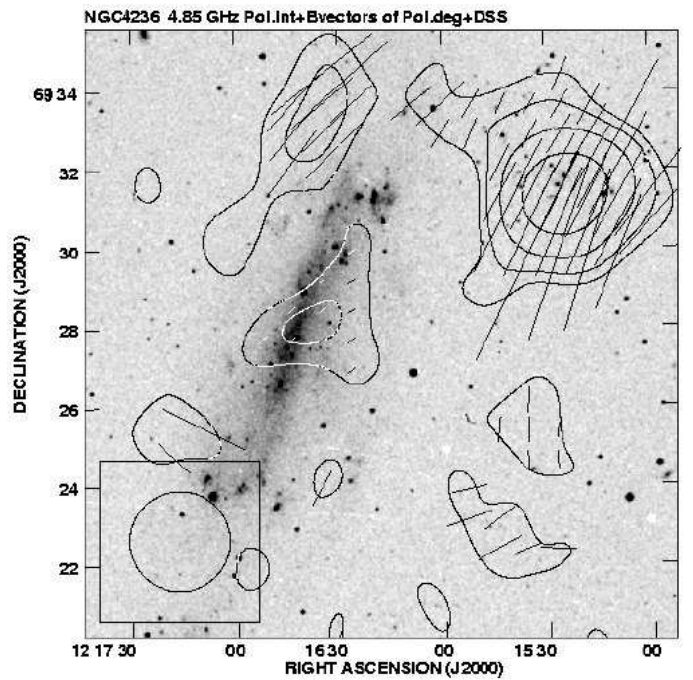}}
\caption{
Contours of polarized intensity and B-vectors of polarization degree of 
NGC\,4236 at 4.85\,GHz superimposed onto the blue image from DSS. The contour 
levels are (2, 4, 8, 16, 32) $\times$ 0.04\,mJy/b.a. A vector of length of 1$'$ 
corresponds to a polarization degree of 6.6\%.  The map resolution is 2\farcm 5~HPBW.
}
\label{4236pi}
\end{figure}

At 1.4\,GHz we used the single horn system in the primary focus of
the Effelsberg 100-m telescope. IC\,2574 was observed in a single band at 
1.400~GHz with a bandwidth of 20~MHz, whereas for NGC\,4256 and NGC\,4656 the 
band was splitted into two independent channels centered on 1.395~GHz and 
1.408~GHz, respectively, with a bandwidth of 14~MHz each. We obtained 
21 good quality coverages for NGC\,4236 and 14 coverages for NGC\,4656 with 
field sizes of $84\arcmin\times84\arcmin$, respectively. For IC\,2574 we got 
12 coverages with a field size of $60\arcmin\times60\arcmin$. The maps were 
scanned in alternating directions along R.A. and Dec. The pixel size and 
separation between the scans were set to $3\arcmin$. 
The scanning velocity was 4\degr/min for NGC\,4236 and NGC\,4656, and 2\degr/min 
for IC2574. We obtained total power maps of the galaxies with an angular 
resolution of 9\farcm 3~HPBW and an r.m.s noise of 5\,mJy/b.a., 7\,mJy/b.a.
and 8\,mJy/b.a. for NGC\,4236, NGC\,4656 and IC\,2574,
respectively.

\begin{figure}
\resizebox{\hsize}{!}{\includegraphics{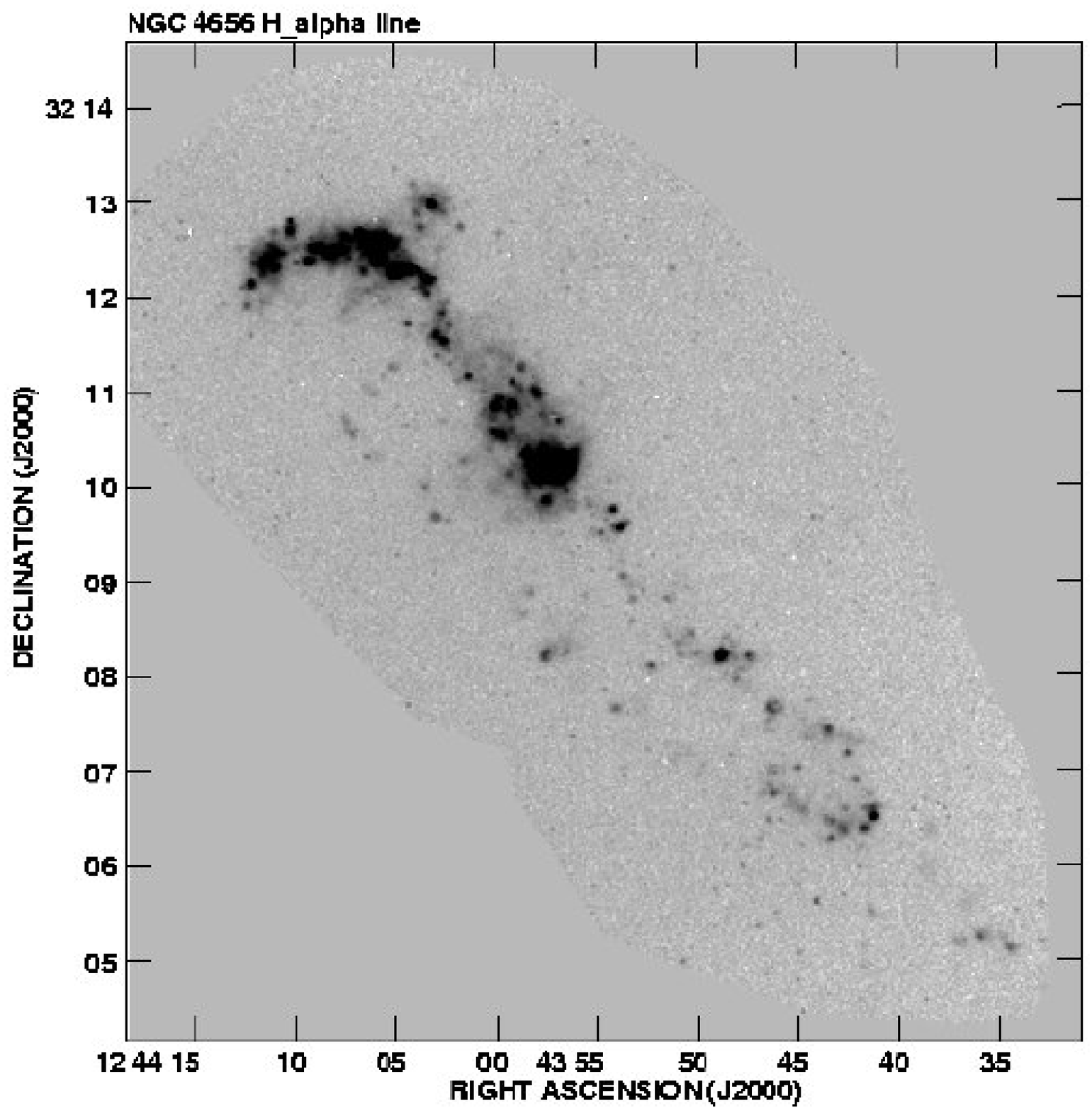}}
\caption
{
The H$\alpha$ image of NGC\,4656 with the optical continuum subtracted
}
\label{4656hal}
\end{figure}

\begin{figure}
\resizebox{\hsize}{!}{\includegraphics{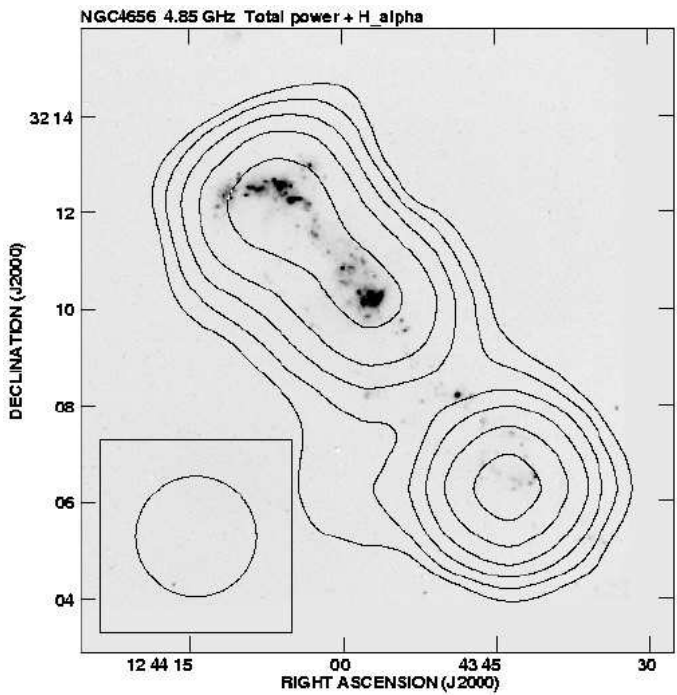}}
\caption
{
Total power contours of NGC\,4656 at 4.85\,GHz superimposed onto the 
H$\alpha$ image. The contour levels are (3, 5, 8, 12, 18, 26)$\times$ 
0.65\,mJy/b.a. The map resolution is 2\farcm5~HPBW.
}
\label{4656tp} 
\end{figure}

\begin{figure}
\resizebox{\hsize}{!}{\includegraphics{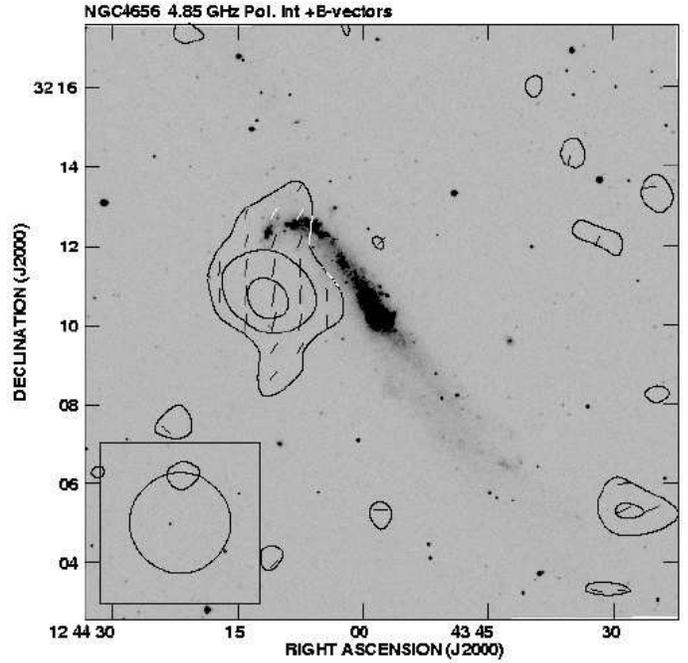}}
\caption{
Map of polarized intensity of NGC\,4656 at 4.85\,GHz with B-vectors of
the same  quantity overlaid upon the DSS red image. The contour levels 
are (2, 3, 4, 8, 16)$\times$ 0.08\,mJy/b.a. A vector of 1\arcmin \, length 
corresponds to a polarized intensity of 0.59 mJy/b.a. The map resolution is 
2\farcm 5~HPBW.
}
\label{4656pi}
\end{figure}

\begin{figure}
\resizebox{\hsize}{!}{\includegraphics{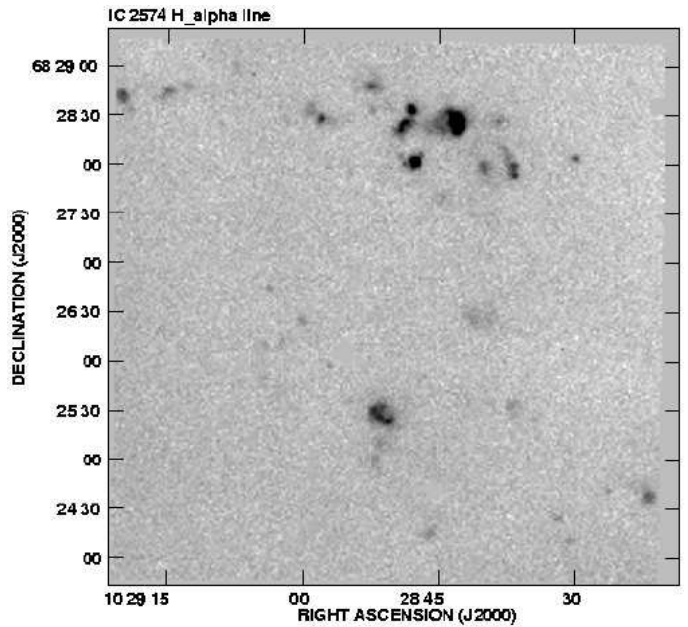}}
\caption{
H$\alpha$ image of IC\,2574 with the optical continuum subtracted. The map
covers the most intense star formation regions in the northern part of the galaxy.
}
\label{2574hal}
\end{figure}

\begin{figure}
\resizebox{\hsize}{!}{\includegraphics{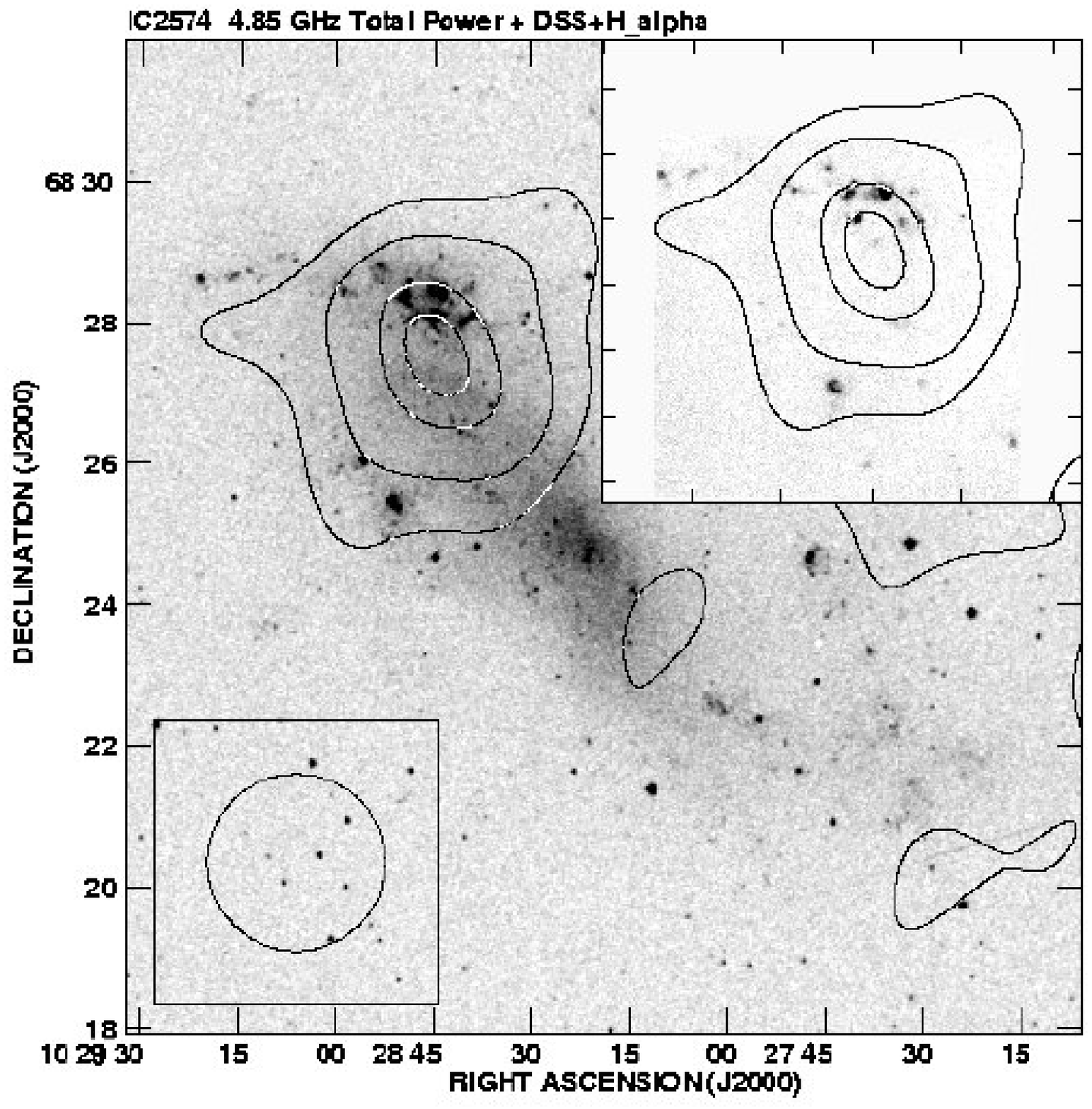}}
\caption{
Contour map of IC\,2574 at 4.85\,GHz superimposed onto the blue DSS
image. The contour levels are (3.2, 7, 12, 14)$\times$ 0.35\,mJy/b.a. The map 
resolution is  2\farcm 5~HPBW. The inserted figure represents the same contour
map overlaid on our H$\alpha$ map covering the northern part of the galaxy.
}
\label{2574tp}
\end{figure}

\section{Observational results}

Below we present the radio data separately for the individual
galaxies. To find possible background sources contributing
to the integrated flux densities we compared our
total power maps at 4.85\,GHz with  high-resolution ones at
1.49\,GHz from Condon (\cite{condon}), NVSS and FIRST. In our 1.4\,GHz data  the
contribution from all point sources not obviously related to the
galaxy was subtracted from our integrated flux densities.
At 4.85\,GHz we applied the ``best subtraction'' (Chy\.zy et al.
\cite{chris_ic}) at the positions of background sources to
estimate their contribution at 4.85\,GHz. These values were then
used to derive the background-free fluxes at this frequency,
listed in Tab.~\ref{fth}.

\subsection{NGC\,4236}
\label{sect:n4236}

NGC\,4236 shows two groups of large \ion{H}{ii} regions at both
ends of the disk   (Fig.~\ref{4236hal}). They are embedded in
diffuse H$\alpha$ emission. The  centre of the galaxy contains
only small \ion{H}{ii} regions and  much less  diffuse emission.
Some of the ionized gas clumps are found as far as 3\arcmin \,
or 3.7\,kpc (inclination corrected)
from the disk plane. This is too far to belong to the disk. Hence, NGC\,4236
clearly possesses extraplanar H$\alpha$ emission from regions up
in the halo.

The radio emission in NGC\,4236 at 4.85\,GHz has a central peak in
the region  where  no large groups of \ion{H}{ii} regions are
found (Fig.~\ref{4236tp}).  The NW cluster of \ion{H}{ii} regions
corresponds to a radio extension while  the SE one is
associated with a separate source. The  gap between this feature 
and the central peak corresponds to a region particularly
deficient in H$\alpha$ emission. 

The polarized sources at  R.A.$_{2000}=\rm 12^h15^m30^s$, 
Dec.$_{2000}=69\degr24\arcmin$ and at  R.A.$_{2000}=\rm
12^h15^m30^s$, Dec.$_{2000}=69\degr34\arcmin$ are composed of
four point sources  visible in Condon's (\cite{condon}) and NVSS 
maps. They are certainly  background objects. They do not
interfere with the galaxy  measurements at 4.85\,GHz but had to be
removed from the 1.4\,GHz  integrated flux. The polarized
extension NE of the northernmost star-forming  clump turned out
to be a background source, as well. It was deleted in our 
integrated flux density
determinations at both frequencies. The only detectable polarized
signal not attributed to background sources comes from the central
region of NGC\,4236  (Fig.~\ref{4236pi}).

We note that Turner \& Ho (\cite{turner}) found three point
sources in the central region of NGC\,4236 which may be radio
supernovae. They  may at least
partly explain the polarization in the central region.

Integration of the total power map in a polygonal region
encompassing the  detectable emission (without background
sources) yields an integrated flux  density of NGC\,4236 at
4.85\,GHz of $23\pm 3$\,mJy. Integration of the map of  polarized
intensity in the same area yields a polarized flux density of 
$0.37\pm 0.13$\,mJy which means 3-sigma detection. The central peak 
gives 13\,mJy/b.a. in total intensity and 0.18\,mJy in polarized 
intensity which means detection at 4.5 sigma confidence level. Integration 
of the 1.4\,GHz map yields a total flux density (corrected for 
background sources) of $48\pm 6$\,mJy. 

\subsection{NGC\,4656}
\label{obs4656}

In NGC\,4656 almost all H$\alpha$ emission is concentrated in the
NE  part of  the disk (Fig.~\ref{4656hal}). This region of the
galaxy is populated by large groups of giant \ion{H}{ii}
regions surrounded by diffuse emission. At the northeastern
end of the disk they make a hook-like appendix, extending  by 
more than 1\arcmin\ to the east. The SW part of the disk shows
very weak  H$\alpha$  emission and very few small \ion{H}{ii}
regions. We  also note a  weak ridge of H$\alpha$ emission
running parallel to the  disk plane at a  distance of about
1\arcmin\ towards the SE. In the NE half of the disk, the 
space  between this ridge and the disk plane is filled with faint
diffuse  spurs. 

The total power radio continuum of NGC\,4656 at 4.85\,GHz 
(Fig.~\ref{4656tp}) shows two basic components: an elongated 
feature  coincident with the NE part of the disk containing
clusters of  large \ion{H}{ii}  regions and a bright unresolved
source in the SW disk located within the optical limits of the
galaxy. This source is found to be a very small object (less than few
arcseconds in size) as its flux density from NVSS (with
50\arcsec~HPBW resolution) and from FIRST (resolution 10 times
better) is almost the same. It is likely to be a very compact
source with no optical counterpart at its  well-measured (arcsec
accuracy) position from FIRST. We assumed it to be a strong 
background source and we dropped it from further analyses.
 
The only obviously polarized feature in NGC\,4656 is the peak of polarized
intensity east of the northern disk boundary, south of the hook-like H$\alpha$
and optical extension  (Fig.~\ref{4656pi}). In this disk region we identified
in the FIRST survey three background sources, coinciding with the polarized
peak. The whole polarization from this region was assumed to be due to
background sources and dropped from further analyses, only the upper limit of
the integrated polarized flux after its subtraction has been estimated. 
Integration of the total power map in the polygonal area encompassing all the
visible radio emission yields an integrated flux density of NGC\,4656 at 
4.85\,GHz of $42\pm 5$\,mJy. Integration of the map of polarized intensity 
over the same area as for the total power power emission yields 
only an upper limit for the polarized flux of 0.44\,mJy at 2-sigma 
confidence level. At 1.4\,GHz we obtain an integrated background-free 
flux density of $92\pm  8$\,mJy.

\subsection{IC\,2574}

The H$\alpha$ emission in IC\,2574 is almost entirely concentrated NE of  the 
optical disk centre (Fig.~\ref{2574hal}). Its distribution corresponds very 
well to the optically bright star-forming regions. There is not much diffuse 
H$\alpha$ emission in our map.

The area around the galaxy has a relatively low confusion level
(unresolved background sources), 
thus we could attain an r.m.s. noise  of 0.4\,mJy/b.a.
The distribution of the radio continuum at 4.85\,GHz in IC\,2574 is 
highly  asymmetric with almost all radio emission concentrated in
the  vicinity of the NE  star-forming clump (Fig.~\ref{2574tp}). 
Down to the mentioned r.m.s. level no emission has been found
from the  main galaxy's body.

In the case of IC\,2574 there was no need to account our integrated flux
densities for a contribution from background sources. Integration of the total
power map in a polygonal area yields an integrated flux density of IC\,2574 at
4.85\,GHz of $9.5\pm 1.4$\,mJy. Integration of the map of polarized
intensity in the same area results only in an upper limit of the polarized flux
density of $0.46$\,mJy at 2-sigma confidence level. At 1.4\,GHz we obtain an
integrated, background-free flux density of $19\pm 8$\,mJy. 

\section{Discussion}
\label{disc}

\subsection{The thermal emission}
\label{thermi}

We determined the thermal emission from the galaxies in our
sample in two ways. First, we determined their radio spectral indices
from the background-free integrated flux densities. Then we used
a standard procedure of  separating thermal and nonthermal
emissions in the way similar to that described  by Chy\.zy et al.
(\cite{chris_ic}). We assumed the nonthermal spectral index
$\alpha_{nt}$ subsequently equal to 0.8, 0.9 or 1.0. 
The results: flux densities S$_{tot}$ at 4.85
and 1.4~GHz, observed spectral index $\alpha_{obs}$, the assumed
nonthermal one $\alpha_{nt}$ as well as radio and optically
derived thermal fluxes S$_{th}$ and thermal fractions f$_{th}$ are
collected in Tab.~\ref{fth}. 

We also determined the thermal flux and the thermal fraction
using the calibrated H$\alpha$ maps in the same way as in Chy\.zy
et al.  (\cite{chris_ic}). To correct the H$\alpha$ fluxes of our
galaxies for absorption we used the total  (galactic and internal)
extinction corrections in B and/or in V-colour as well as  the
reddening E$_{B-V}$ from the LEDA database. They are computed
using the most  recent data also including corrections for local
anomalies. In converting them  to the absorption in the red
domain we compared several methods. First, we converted
independently the E$_B$ and E$_V$ into E$_R$ using the
appropriate ``dust colours''  from the NED database (assuming the
colour of  total absorption similar to that in our Galaxy).
Alternatively we scaled E$_B$   and E$_V$ to E$_R$ using the
extinction ratios from Schlegel et al.  (\cite{schleg}).
Additionally, we applied the statistical analyses of absorption
by Cardelli et al. (\cite{cardel}) to E$_{B-V}$ taken from LEDA.
The above  variety of methods yielded
thermal fractions in the range 0.45 -- 0.63 (NGC  4236), 0.5 -- 0.9
(NGC 4656) and 0.62 -- 0.78 for IC 2574. The mean values  averaged
over all methods, together with r.m.s. method-to-method scatter,
are  given in Tab.~\ref{fth}.

Niklas et al. (\cite{niklas2}) gave the mean thermal fraction at
10.45\,GHz f$_{th} = 30\%\pm5\%$ with 15\% of the galaxies in
their sample of 74 galaxies having f$_{th} \ge 45\%$,
independent of morphological type. They found a mean nonthermal
spectral slope $\alpha_{nt}$ = 0.83 for their sample of spiral
galaxies. This implies a mean f$_{th} \simeq 24\% $ at 4.85\,GHz.
Soida et al. (\cite{soida3627}, \cite{4254}) simulated the radio
spectra composed of assumed fractions of thermal and nonthermal
emission with the slope of the latter adjusted to obtain the
observed spectral index. They conclude that for the rapidly
star-forming spirals NGC\,3627 and NGC\,4254  the model spectrum
poorly fits  observations if f$_{th}$ at 10.55\,GHz is higher
than 40\%. This implies thermal fractions of these objects at
4.85\,GHz  considerably lower than 25\%.

The uncertainties of both, the radio and optical method are quite
large, because  of a poor knowledge of nonthermal spectra and
optical extinction. We note, however, that for two objects:
NGC\,4656 and IC 2574 the {\em lowest } thermal fractions obtained
for various estimates of optical extinction are comparable
to the radio-derived f$_{th}$ for the {\em steepest nonthermal
spectra. In NGC\,4236 the ranges of the optically and
radio-determined f$_{th}$ overlap, the optical ones also being  somewhat
higher. All this seems to favourize rather steep nonthermal
spectra. Both the radio-derived thermal fractions assuming
$\alpha_{nt}\ge 0.9$ and lowest values from the H$\alpha$ line
are consistently higher by a factor of $\ge 1.9$ than the mean for
radio-bright spirals studied by Niklas (\cite{niklas2})}. To make
our estimates  similar to those for normal spirals we would need
to assume simultaneously that the  nonthermal spectrum has a
slope of 0.65 -- 0.70 (considerably flatter than for normal
spirals) and that all our absorption estimates are too high by a
factor of two. We consider such a coincidence as rather unlikely.
It seems that the radio emission of our galaxies really shows  a
considerably increased thermal content compared to normal
galaxies. The radio supernovae in NGC\,4236 contribute some 8\,mJy
to the total radio flux at 4.85\,GHz being themselves
predominantly nonthermal point sources. Thus the thermal fraction
of the {\em diffuse} emission from NGC\,4236 may be even higher
than given in Tab.~\ref{fth}.

\begin{table*}
\begin{center}
\caption[]{Confusion-corrected fluxes, spectral indices,
   observed and assumed nonthermal, as well as thermal fluxes
   and fractions of observed galaxies at 4.85\,GHz}
\begin{tabular}{lrrcrrrrr}
\hline
Galaxy  &     S$_{tot}$&S$_{tot}$& $\alpha_{obs}$&$\alpha_{nt}$& S$_{th}$(4.85) &  f$_{th}$(4.85)
& S$_{th}$(4.85)&
f$_{th}$(4.85)\\
         &4.85&1.4 &   &       &  radio & radio   &  opt         & opt       \\
         &[mJy]&[mJy]& &       &  [mJy] &  &[mJy]                &           \\
\hline
NGC\,4236 & 23& 48 &0.59$\pm 0.15$&  0.8  &  9.0   &  0.39   &   11.7        &
0.51$\pm 0.07$      \\
         &   &    &    &  0.9  &  11.6  &  0.50   &              &           \\
         &   &    &    &  1.0  &  13.6  &  0.59   &              &           \\
\\
NGC\,4656 & 42& 92 &0.63$\pm 0.12$&  0.8  &  13.7  &  0.33   &  26.9        &
0.64$\pm 0.14$       \\
         &   &    &    &  0.9  &  18.9  &  0.45   &              &            \\
         &   &    &    &  1.0  &  22.9  &  0.55   &              &            \\
\\
IC\,2574  &9.5&19  &0.56$\pm 0.36$&  0.8  &  4.2   &  0.45   &   6.5       &
0.68$\pm 0.06$      \\
         &   &    &     &  0.9  &  5.2   &  0.55   &              &            \\
         &   &    &     &  1.0  &  6.0   &  0.63   &              &            \\
\hline
\label{fth}
\end{tabular}
\par
{\it Note: the uncertainties of optically derived f$_{th}$
result from various estimates of total extinction}
\end{center}
\end{table*}

The highest thermal fraction in a spiral galaxy so far has been
found in the edge-on spiral NGC\,5907, i.e. 54\% at 4.85\,GHz
(Dumke et al. \cite{dumke00}). This galaxy has a rather low star
formation rate and Dumke et al. (\cite{dumke00}) proposed that
the high thermal fraction may not be due to an increased thermal
emission but rather to a {\em deficiency of nonthermal emission}.
(see Sect.~\ref{magnetic}).

\subsection{Distribution of thermal and nonthermal emission}

\begin{figure}
\resizebox{\hsize}{!}{\includegraphics{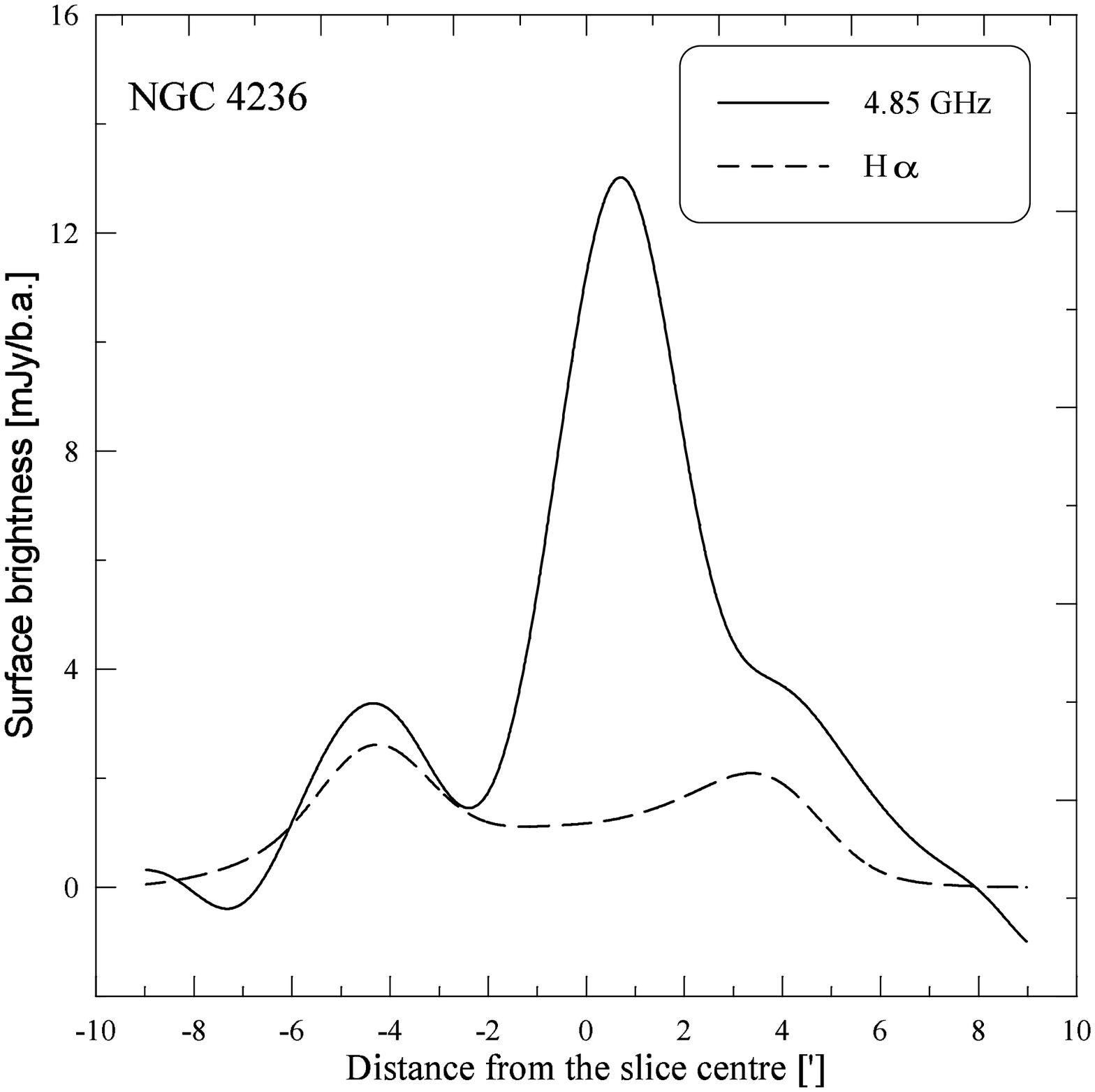}}
\caption{
Brightness profiles of the radio emission at 4.85\,GHz and of
H$\alpha$  emission along the disk plane of NGC\,4236. The slice
centre is at  R.A.$_{2000}=\rm 12^{h}16^{m}39\fs 7$
Dec.$_{2000}=+69\degr 27\arcmin  41\arcsec$ and the position angle
is 334\fdg6. The H$\alpha$ image is  convolved to the
resolution of the radio map of 2\farcm 5~HPBW. }
\label{4236sli}
\end{figure}

\begin{figure}
\resizebox{\hsize}{!}{\includegraphics{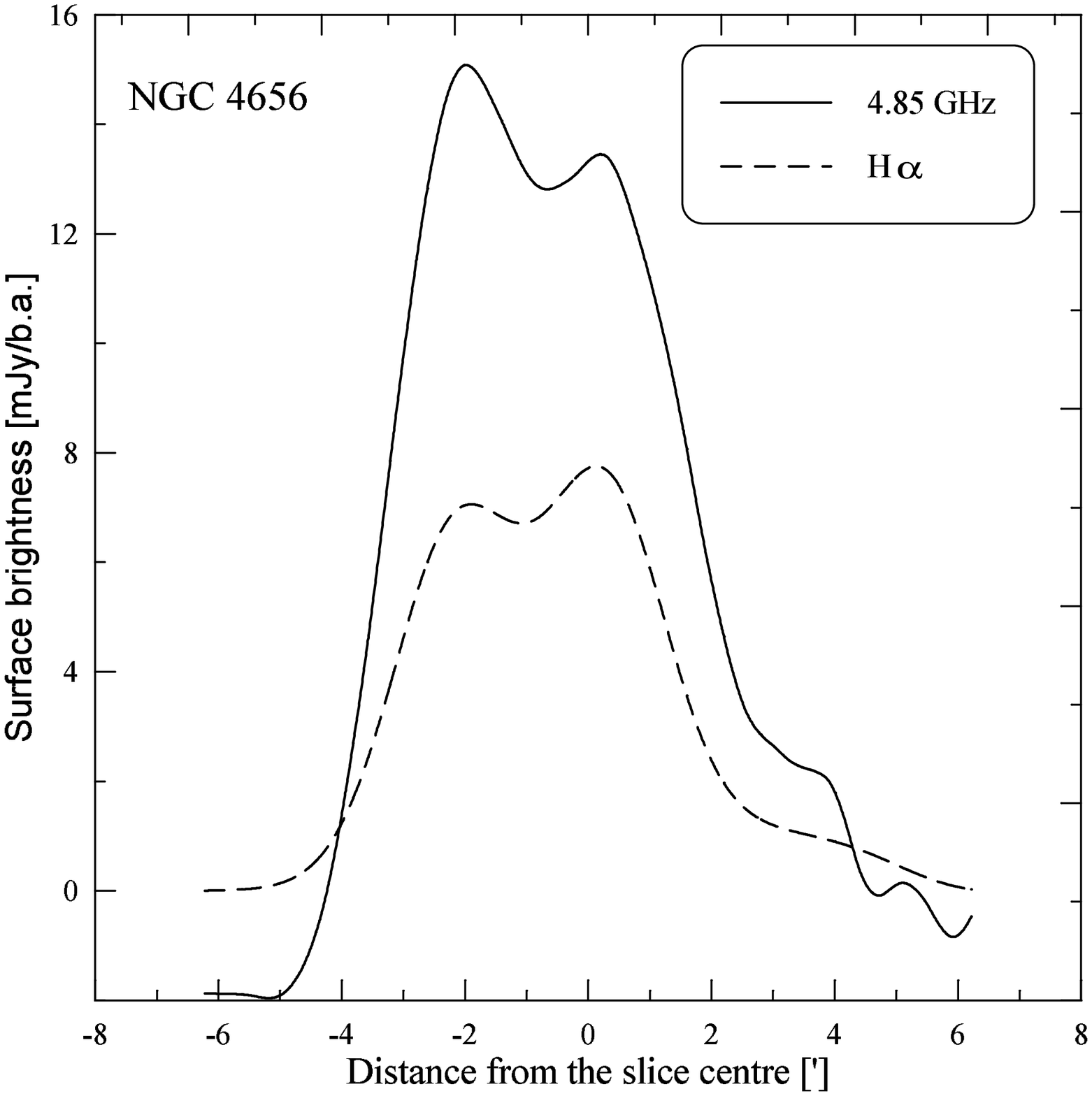}}
\caption{
Brightness profiles of the radio emission at 4.85\,GHz and of
H$\alpha$  emission along the disk plane of NGC\,4656. The slice centre is at
R.A.$_{2000}=\rm 12^{h}43^{m}58\fs 9$ Dec.$_{2000}=+32\degr 10\arcmin
32\arcsec$ and the position angle is 221\degr. The H$\alpha$ image is
convolved to the resolution of the radio map of 2\farcm 5~HPBW.
}
\label{4656sli}
\end{figure}

Two of our galaxies (NGC\,4236 and NGC\,4656) possess clearly
defined disks. For them we constructed profiles of total radio
(4.85\,GHz) and thermal emission (derived from the H$\alpha$ line)
along the major axis. The H$\alpha$ data were smoothed to the
resolution of the radio maps (2\farcm 5). The profiles are centered
on R.A.$_{2000}=\rm 12^{h} 16^{m} 39\fs 7$ Dec.$_{2000}=+69\degr
27\arcmin  41\arcsec$ and R.A.$_{2000}=\rm 12^{h} 43^{m} 58\fs 9$
Dec.$_{2000}=+32\degr  10\arcmin 32\arcsec$, respectively. A
strong point source in the southern part of  NGC\,4656 has been
subtracted. The profiles  are shown in Figs.~\ref{4236sli} and
\ref{4656sli}.

In NGC\,4656 the distribution of  total and thermal emission look
similar (Fig.~\ref{4656sli}), with  some minor asymmetry between
the peaks. Along the whole profile the thermal  fraction seems to
undergo only small variations.

NGC\,4236 looks different in this respect (Fig.~\ref{4236sli}). A
strong,  unresolved central radio peak has no counterpart in the
H$\alpha$ emission. Its amplitude can be well accounted for by
the ``radio supernovae'' found there by Turner \& Ho
(\cite{turner}). In the remaining disk (e.g. at the distance of
5\arcmin \, in Fig.~\ref{4236sli}) the thermal fraction may be
very high, reaching some 70\%. This supports our conjecture from
Sect.~\ref{thermi} that NGC\,4236 may be dominated by thermal
emission, having very weak global magnetic fields. We conclude
that mechanisms producing the total  magnetic fields in the
presence of weak star formation are rather inefficient.

\subsection{Integrated properties}

\begin{figure}
\resizebox{\hsize}{!}{\includegraphics{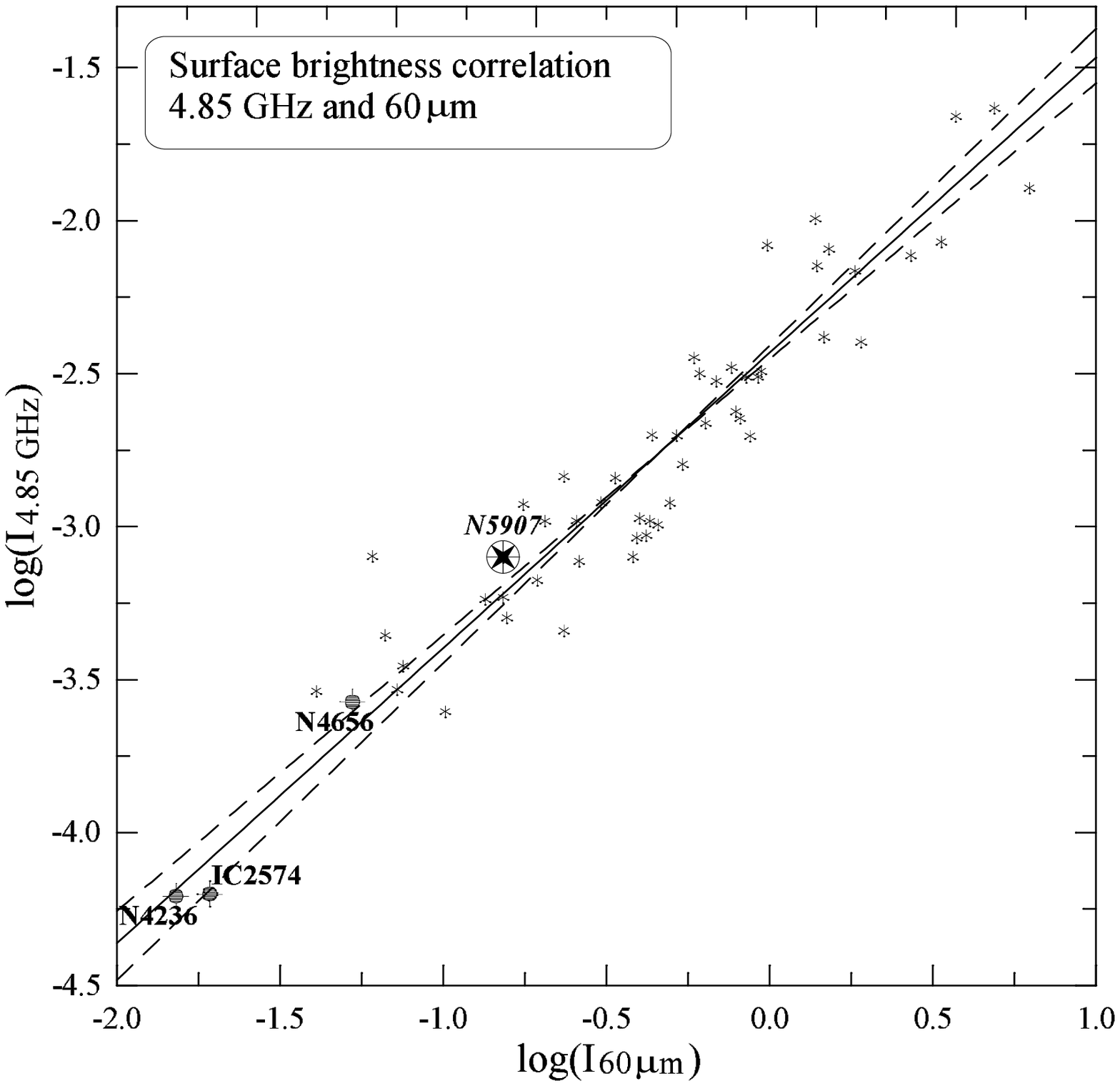}}
\caption{
Positions of the galaxies studied on the correlation diagram
between the face-on corrected surface brightness at 60$\mu$m and
at 4.85\,GHz (in Jy/arcmin$^2$). Small stars show  the normally star-forming
galaxies from the sample by Gioia et al. (1982), crossed dots --
our objects. The solid line shows the orthogonal fit (slope 0.96)
to objects from Gioia et  al. (1982), dashed lines (slopes of
0.90 and 1.04) are the lower and upper limits to the best-fit slope
(see text). A  thick, encircled cross marks the position of
NGC\,5907, compared in the text to our galaxies.
}
\label{corel}
\end{figure}

Fig.~\ref{corel} shows the positions of our galaxies on a
radio/FIR correlation between the surface brightness  at
60\,$\mu$m and at 4.85\,GHz derived for a sample of normal
spirals, measured at this frequency by Gioia et al. (\cite{ggk},
GGK). The fluxes at 60\,$\mu$m are taken from Helou \& Walker
(\cite{iras1}) and if  not present there from Moshir et
al.~(\cite{iras2}). The values of FIR and radio mean surface
brightness have been obtained by dividing the corresponding
integrated flux  densities (in Jy) by the area within the
extinction and inclination-corrected galaxy radius in the B band
at 25mag/arcsec$^2$ taken from the LEDA database. No correction
for thermal  emission has been made at this stage. As both the
radio  and FIR brightness are random variables we used the
orthogonal  fit to determine the slope of the best-fit line. In
such case  simple regressions of FIR vs. radio and radio vs. FIR
give the  minimum and maximum slope allowed by the data scatter,
forming  so-called ``regression scissors''. We used them as lower
and upper limits to the slope of the best-fit line.

Our objects significantly extend the radio/FIR correlation at
4.85\,GHz towards  the low surface brightness region. They follow
well the correlation line with a slope of $0.96\pm 0.06$
derived  from the GGK sample alone. This means that the
total (i.e. thermal and nonthermal) emission of our galaxies of
low surface brightness  follow the same correlation law as
normal star-forming, radio and FIR-bright spirals.

It is difficult to say whether the galaxies would lie below the
correlation line for bright galaxies if only the {\em nonthermal}
emission were considered. This needs a larger sample of galaxies
spanning a broad range of star-forming activities, with radio
spectra and H$\alpha$ data analyzed in a homogeneous way. Such a
systematic study is under consideration.

The mean degrees of nonthermal polarization (or their upper
limits) of all galaxies of our sample are given in
Tab.~\ref{mfield}. We compared these rather low values with
those of other highly-inclined spiral galaxies of similar angular
size measured with the same angular resolution, hence influenced
by beam depolarization in a similar way. These objects
were observed at 4.85\,GHz with the 100-m radio telescope by one
of us (MK) in other works. Our comparison  spirals:
NGC\,891, NGC\,3628, NGC\,4565  NGC\,5907  and  NGC\,4631 have {\em
nonthermal polarization degrees of 3.2\%, 5.3\%, 13.7\%, 3.2\%
and 4.9\%, respectively,} assuming a mean thermal fraction for
spirals of 20\% at 4.85\,GHz (see Sect.~\ref{thermi}). Adopting
for NGC\,4565 and NGC\,5907, which have relatively low star
formation, thermal fractions as high as 60\% at 10.45\,GHz
(Niklas et al.~\cite{niklas2}) which yields  40\% at  4.85\,GHz,
we obtain their {\em  nonthermal} degrees of polarization
$\simeq$ 17\% and 4\%, respectively. The mean degree of
polarization in NGC\,4236 and the upper limit for NGC\,4656
(Tab.~\ref{mfield}) are below all these values. In NGC\,4236
the polarization degree may be even lower if the bulk of
polarization is due to the radio supernovae (see
Sect.~\ref{sect:n4236}). 
We suggest that these two galaxies may have
less ordered magnetic fields (hence regular fields weaker
compared to total ones) than other more rapidly star-forming
edge-on spirals. This means that also the mechanism producing
regular magnetic fields may be less efficient in weakly
star-forming galaxies. We note, however,
that in case of IC\,2574, a weak total power emission leads to a
yet inconclusive upper limit for the polarization degree.

\subsection{Magnetic fields}
\label{magnetic}

\begin{table}
\caption[]{Magnetic field strengths and degree of nonthermal
polarization in the observed galaxies.}
\begin{center}
\begin{tabular}{lrrr}
\hline
Galaxy     &  B$_{t}$  &    B$_{reg}$  &  Pol. degree\\
         & [$\mu$G] &  [$\mu$G] &   [\%] \\
\hline
NGC\,4236 total &   $4.4 \pm 1.1$     & $0.6 \pm 0.2$  & 1.7$\pm 1.0$   \\
NGC\,4236 5\%   &   $4.7 \pm 1.1$     & $0.7 \pm 0.2$  & 1.6$\pm 0.9$   \\
\\
NGC\,4656 total &   $4.7 \pm 1.1$     & $<0.9 \pm 0.3$ & $<2.1^{\rm a}$ \\
NGC\,4656 2 r.m.s. &$4.9 \pm 1.2$     & $<1.0 \pm 0.3$ & $<2.0^{\rm a}$ \\
\\
IC\,2574 total  &   $4.0 \pm 1.1$     & $<1.2 \pm 0.3$ &  $<9.3^{\rm a}$   \\
IC\,2574 2 r.m.s.  &$4.3 \pm 1.1$     & $<1.3 \pm 0.3$ &  $<8.4^{\rm a}$   \\
\hline
\label{mfield}
\end{tabular}
\end{center}
$^{\rm a}$ upper limit at the $2\sigma$ level of integrated flux density 
uncertainty
\end{table}

We computed the equipartition magnetic field strengths in our
galaxies from the nonthermal surface brightnesses (according to
Beck \& Krause \cite{beck05}) in two ways. First, we used the
mean value of the nonthermal surface brightness at 4.85\,GHz
integrated over the whole extent of radio emission or (in case of
IC\,2574) over the optical extent. As a second approach, we took
the mean nonthermal emission within the area delineated by 5\% of
the maximum signal or the area with emission above 2 r.m.s. map
level,  taking the larger of these thresholds. We assumed a
proton-to-electron energy ratio of 100. For each galaxy we used the
nonthermal spectral index that yielded the best agreement between
the radio and optically  determined thermal fraction in
Tab.~\ref{fth}. A synchrotron face-on disk thickness of 1\,kpc
was assumed.  We determined the average brightness of the radio
emitting region in IC\,2574 by taking it roughly spherical with a
diameter of 4.5\,kpc. The resulting total and regular magnetic
field strengths as well as the mean degree of {\em nonthermal}
polarization (corrected for polarized background sources) are
shown in Tab.~\ref{mfield}. Their errors include the 50\%
uncertainties of the above parameter values.

Our galaxies have relatively low total mean magnetic field
strengths (in the range 4.0 to 4.9$\,\mu$G) when compared to 
normal spirals (for which the mean value depending on the
galaxy sample is $8-10\,\mu$G; Beck et al. \cite{beck96}). 
In NGC\,4236 the mean total ``diffuse'' magnetic field may be even
weaker if the ``radio supernovae'' discussed by Turner \& Ho
(\cite{turner}) and in Sect.~\ref{thermi} are removed.

The galaxies of our sample belong to objects with the weakest
total fields so far measured in galaxies. Other spirals with weak
total and regular magnetic fields are e.g. NGC\,5907 (Dumke et al.
\cite{dumke00}) and the Sombrero galaxy M104 (Krause et al.
\cite{krause}). Although the former is a late type spiral and the
latter an early-type spiral, both have low star formation rates.
Furthermore, like the galaxies of our sample, NGC\,5907 has a high
thermal fraction which may be due to a deficiency of synchrotron
emission, which in turn indicates a low magnetic field strength.
A similar deficiency of synchrotron emission seems to be present in
galaxies studied in this work.

According to Dumke et al. (\cite{dumke00}) a higher star
formation activity may lead to a much more significant increase
of the number density of relativistic electrons and a more
effective amplification of the magnetic fields, and therefore to
a higher fraction of nonthermal  (synchrotron) radiation than for
galaxies with a low star-formation rate (SFR). A more efficient
magnetic field generation is predicted e.g. by the model of the
dynamo  process driven by Parker-type magnetic instabilities
boosted by star-formation (Hanasz et al. \cite{hanasz04}). 
The pressure of cosmic rays produced by supernovae accelerates
the instabilities which in turn leads to a rapid amplification
of both total and regular magnetic fields (the latter by fast
reconnection). This scenario would be consistent with the
situation in the galaxies in our sample: they all have a high
thermal fraction, weak total emission, and low star formation
activity. Indeed, Niklas et al. (\cite{niklasphd}) found that the
lowest thermal fractions occur in galaxies with the highest SFR.

Niklas \& Beck (\cite{niklas3}) explained the radio--FIR
correlation by a simple model where the energies of turbulent gas
motions, magnetic fields and cosmic rays are in equipartition.
As a result, the field strength $B_t$ increases as $B_t\propto
SFR^{1/(2N)}$ where $N$ is the exponent of the Schmidt law
($SFR\propto\rho^N$, Schmidt, \cite{schmidt2}), and the
nonthermal surface brightness $I_{nth}$ increases as
$I_{nth}\propto SFR^{~(3+\alpha_{nt})/(2N)}$. 
For $N\simeq1.4$ and $\alpha_{nt}=0.9$ we get
$I_{nth}\propto SFR^{~\approx 1.4}$. As the thermal radio surface
brightness is expected to increase linearly with $SFR$, 
the ratio of nonthermal to thermal emission increases
with $SFR^{~\approx 0.4}$. 
This simple model can basically explain our result that 
the nonthermal radio emission and the ratio of nonthermal
to thermal emission are
\emph{nonlinear} functions of the star-formation rate.  Below
some threshold in star-formation activity, the dynamo  may not
work efficiently anymore, and the nonthermal emission may drop
further. Below some low value of $SFR$, the radio emission may be
almost completely thermal, but the radio--FIR correlation is
still valid. A detailed study of these correlations is in
preparation.

\section{Summary and conclusions}

We observed the three angularly large, late-type, slowly
star-forming nearby galaxies  NGC\,4236, NGC\,4656 and IC\,2574
with the Effelsberg radio telescope at  4.85\,GHz in order to
check whether they still host widespread magnetic fields.
Observations at 1.4\,GHz were also made to determine the radio
spectra. Our  results were carefully accounted  for possible
background sources. We also  observed  these galaxies
in the H$\alpha$ line to study the distribution of recent  star
formation and to estimate thermal fractions independent of the
radio  spectrum. The most important results are as follows:

\begin{itemize}

\item[--] The galaxies possess a very clumpy distribution of the
H$\alpha$-emitting gas. The \ion{H}{ii} regions are very
non-uniformly distributed being concentrated in selected parts of
the galaxies in a very asymmetric way. NGC\,4236 shows even some
extraplanar H$\alpha$ emission.

\item[--] A reasonable agreement between mean thermal fractions
determined from the radio  spectrum and from the
extinction-corrected H$\alpha$ line was attained.  The galaxies
show thermal fractions at 4.85\,GHz considerably higher than most
normal spirals, reaching 0.5 -- 0.7 (depending on the assumptions).
This effect is most pronounced for the radio-weakest objects.

\item[--] The galaxies clearly show diffuse nonthermal emission
indicative of widespread magnetic fields. Their mean total surface
radio brightness is  however  smaller than in most normally
star-forming intermediate-type spirals.

\item[--] The fraction of nonthermal
emission in our objects is smaller than for  normally
star-forming spirals which suggests that magnetic fields are even
weaker than one could expect from their low radio brightness.
Their total magnetic fields have mean strengths of 4.0 -- 4.9 $\mu$G.
This result can be explained in terms of the equipartition model
between turbulent gas motions, magnetic fields and cosmic rays,
in which the nonthermal emission increases faster than linearly
with the SFR ($\propto SFR^{~\approx 1.4}$) and that the ratio of nonthermal 
to thermal emission increases with $SFR^{~\approx 0.4}$.

\item[--]The galaxies still fall on the radio-FIR correlation for the
total surface brightness, following the same slope
of $0.96\pm 0.06$ as radio-bright spirals.
Our sample extends the correlation to the lowest values observed
so far.

\item[--] Weak polarization at 4.85\,GHz attributable to the galaxy
has been detected only in NGC\,4236. Even in this case it can be
due to few young supernova remnants. Two galaxies (NGC 4236 and NGC 4656)
have a nonthermal polarization degree below values
measured for other edge-on spirals at 4.85\,GHz with a similar
resolution. The degree of magnetic field regularity of these two
weakly star-forming spirals is lower than that of normally 
star-forming and rapidly rotating edge-on spirals. The third 
object (IC 2574) is too weak for any definite constraints.
\end{itemize}

In this work we demonstrated that the slowly
rotating and weakly star-forming spirals observed by us may
possess widespread magnetic fields. The field  generation
mechanism, especially that of the regular magnetic field, seems
to be less efficient in these objects than in rapidly
star-forming spirals. This gives arguments  for magnetic field
generation theories involving unstable processes (like  Parker
instabilities) energized by star-forming processes. 
To what extent these results apply to slowly
star-forming galaxies in general will be a subject of future
projects dealing with a larger sample of objects. 

\begin{acknowledgements}

The authors, (K.Ch., M.S. and M.U.) are indebted to Professor
Richard  Wielebinski from the Max-Planck-Institut f\"ur
Radioastronomie (MPIfR) in Bonn for the invitations to stay at
this institute where substantial parts of this  work were done. A
large part of the work has been done in the framework of the
exchange program between the Jagiellonian University and
Ruhr-Universit\"at  Bochum. We are grateful to Dr. Elly
Berkhuijsen for valuable comments and careful reading of the manuscript.
We also would like to thank an anonymous referee for his 
comments. We acknowledge the usage of the HyperLeda database
(http://leda.univ-lyon1.fr), the NRAO
VLA Sky Survey (NVSS) and Faint Images of the Radio Sky at Twenty-cm
(FIRST) survey. This work was supported by a grant no.
PB0249/P03/2001/21 from the Polish Research Committee (KBN).
\end{acknowledgements}

\end{document}